\documentclass{PoS}

\PoS{PoS(LAT2005)182}

\usepackage{epsfig}

\newcommand{\kv}{{\mathbf k}}
\newcommand{\pv}{{\mathbf p}}
\newcommand{\xv}{{\mathbf x}}
\newcommand{\vecnul}{{\mathbf 0}}
\newcommand{\om}{\omega}
\newcommand{\be}{\begin{equation}}
\newcommand{\ee}{\end{equation}}
\newcommand{\bra}{\langle}
\newcommand{\ket}{\rangle}
\newcommand{\id}{1\!\!\!1}

\newcommand{\im}{{\mathrm{Im}}}

\newcommand{\tr}{{\mathrm{tr}}}

\title{On meson spectral functions at high temperature and nonzero 
momentum}

\ShortTitle{On meson spectral functions at high temperature and nonzero 
momentum}

\author{
\speaker{Gert Aarts}\addtocounter{footnote}{1}\footnote{PPARC Advanced 
Fellow}, 
	Simon Hands\footnote{PPARC Senior Research Fellow} \\
        University of Wales Swansea\\
        E-mail: \email{g.aarts@swan.ac.uk}, \email{s.hands@swan.ac.uk}
	}

\author{Seyong Kim\\
        Sejong University, Seoul \\
        E-mail: \email{skim@sejong.ac.kr}
	}

\author{Jose M.\ Mart{\'i}nez Resco\\
        Brandon University \\
        E-mail: \email{martinezrescoj@brandonu.ca}
	}

\abstract{
 In the high-temperature phase of QCD meson spectral functions at nonzero 
momentum are expected to have a nontrivial and interesting structure. In 
order to provide a reference point for lattice studies employing e.g.\ the 
Maximal Entropy Method, we discuss several characteristics of meson 
spectral functions in the infinite-temperature limit.
 We report on ongoing work in quenched QCD with staggered fermions.
 }

\FullConference{XXIIIrd International Symposium on Lattice Field Theory\\
                 25-30 July 2005\\
                 Trinity College, Dublin, Ireland}

\begin{document}

\section{Spectral functions}

The recent experimental progress in the recreation of the quark gluon 
plasma in relativistic heavy ion collisions (see e.g.\ 
\cite{Adcox:2004mh}) has been an important stimulus for lattice studies of 
spectral functions in the high-temperature phase of QCD.
 Topics discussed so far (ordered according to increasing difficulty) 
include the persistent presence of charmonium bound states above the 
deconfinement transition \cite{Asakawa:2003re,Datta:2003ww,Umeda:2002vr}, 
the thermal dilepton rate \cite{Karsch:2001uw}, and transport 
coefficients, such as the electrical conductivity \cite{Gupta:2003zh} and 
the shear viscosity \cite{Nakamura:2004sy}. In spectral function 
calculations soft energies $\om \lesssim T$ are of particular interest, 
since this is where one expects e.g.\ nonperturbative medium effects and 
collective excitations, predicted by perturbative studies of hot QCD. 
Moreover, transport coefficients follow from current-current spectral 
functions in the limit of vanishing energy (Kubo relations). So far 
numerical results for the soft energy region in the high-temperature phase 
are rather scarce. For the dilepton rate, it was found 
\cite{Karsch:2001uw} (surprisingly) that the spectral function in the 
vector channel $\rho_{\rm V}(\om,\vecnul)$ is consistent with zero when 
$\om\lesssim 2T$. For the electrical conductivity, 
$\rho^{ii}(\om,\vecnul)$ was reconstructed at low energies 
\cite{Gupta:2003zh} and a structure resembling analytical expectations 
\cite{Aarts:2002cc} was obtained.

However, numerically determined euclidean correlators $G_H(\tau,\pv)$ do 
not easily provide knowledge of spectral functions at soft energies. In 
fact, they are largely insensitive to details of $\rho_H(\om,\pv)$ in this 
region because the kernel $K$, which relates $G_H$ and $\rho_H$ via
 \be
\label{eqGrho}
G_H(\tau,\pv) = \int_0^\infty \frac{d\om}{2\pi}\,
K(\tau,\om)\rho_H(\om,\pv),
\;\;\;\;\;\;\;\;\;\;\;\;
K(\tau,\om) = \frac{\cosh[\om(\tau-1/2T)]}{\sinh(\om/2T)},
\ee
 becomes independent of $\tau$ for smaller energies ($K(\tau,\om) \sim 
2T/\om$ when $\om\ll T$). It follows in particular that lattice 
correlators are remarkably insensitive to transport coefficients 
\cite{Aarts:2002cc} (see ref.\ \cite{Petreczky:2005nh} for a recent 
analysis reaching the same conclusion). In view of the experimental 
results on the strongly interacting QGP, suggesting e.g.\ a surprisingly 
small shear viscosity, it is important to gain more experience in the 
reconstruction of spectral functions at energies $\om\lesssim T$ from 
lattice QCD.

Good candidates for such studies are meson spectral functions at nonzero 
momentum $p$ \cite{Aarts:2005hg}. Because of the presence of the lightcone 
at $\om=p$ and the nontrivial spectral weight below the lightcone, 
commonly referred to as Landau damping and due to the scattering of quarks 
with offshell gauge bosons, they are expected to have an interesting 
structure at soft energies. By choosing the nonzero momentum $p$ of the 
order of $T$ or larger, the region below the lightcone can be made 
sufficiently large and (part of) the difficulties present for transport 
coefficients, inherent in the strict limit $\om\to 0$, can be 
circumvented.
 In order to provide a reference point for such calculations, we studied 
meson spectral functions at nonzero momentum in the infinite temperature 
limit, both in the continuum and on the lattice for Wilson and staggered 
(naive) fermions \cite{Aarts:2005hg}. A similar study at zero momentum can 
be found in ref.\ \cite{Karsch:2003wy}. Below we summarize some of our 
findings. In the outlook we report on ongoing work in quenched QCD with 
staggered fermions.

\section{Infinite temperature}

We consider meson spectral functions $\rho_H(t,\xv)=\bra [J_H(t,\xv), 
J_H^\dagger(0,\vecnul]\ket$ at leading order in the loop expansion. The 
currents in the various channels are given by $J_H(t,\xv)=\bar 
q(t,\xv)\Gamma_H q(t,\xv)$ with $\Gamma_H=\{\id, \gamma_5, \gamma^\mu, 
\gamma^\mu\gamma_5\}$. In the continuum the one-loop integral \be 
\rho_H(\om,\pv) = -2\im\,\, T\sum_{l\in \mathbb{Z}} \int 
\frac{d^3k}{(2\pi)^3}\,\tr\,\, S(i\tilde\om_l,\kv)\Gamma_H 
S(i\om_n+i\tilde\om_l,\pv+\kv)\gamma^0\Gamma_H^\dagger\gamma^0 
\Big|_{i\om_n\to\om+i0^+}, \ee where $S(i\tilde\om_l,\kv)$ is the fermion 
propagator, can be done analytically, for arbitrary external momentum $p$ 
and fermion mass $m$. Since the outcome is rather lengthy, we refer to 
ref.\ \cite{Aarts:2005hg} for detailed expressions.
 On the lattice completely analytical results are not available, but 
simple expressions which are easily evaluated numerically can be given. In 
order to do so, we use a 'mixed' representation \cite{Carpenter:1984dd} 
and write the fermion propagator as $S(\tau, \kv) = \gamma_4 S_4(\tau, 
\kv) + \sum_{i=1}^3 \gamma_i S_i(\tau, \kv) +\id S_u(\tau, \kv)$. For 
Wilson fermions, the $S(\tau, \kv)$ functions depend on $\tau$ via 
$\sinh(\tilde \tau E_\kv)$ and $\cosh(\tilde \tau E_\kv)$, where 
$\tilde\tau=\tau-1/2T$, which makes it straightforward to arrive at the 
relationship (\ref{eqGrho}), also on a finite lattice. The resulting 
spectral functions are then given by a sum over the spatial lattice 
momenta, which can be done numerically.

 For naive fermions there are a few changes, due to the time doublers. The 
hyperbolic functions are now multiplied with the staggering factors
$1\pm (-1)^{\tau/a_\tau}$, such that relation (\ref{eqGrho}) is modified 
to
 \be 
\label{eqGrhostag} G_H(\tau,\pv) = 2\int_0^\infty \frac{d\om}{2\pi}\, 
K(\tau,\om)\left[ \rho_H(\om,\pv) - (-1)^{\tau/a_\tau} \tilde 
\rho_H(\om,\pv) \right], 
\ee 
 where $K$ is again the same kernel as in the continuum. The staggered 
partner $\tilde \rho_H$ is related to $\rho_H$ via the replacement 
$\Gamma_H\to \tilde\Gamma_H=\gamma_4\gamma_5\Gamma_H$.

\begin{figure}[b]
 \centerline{\epsfig{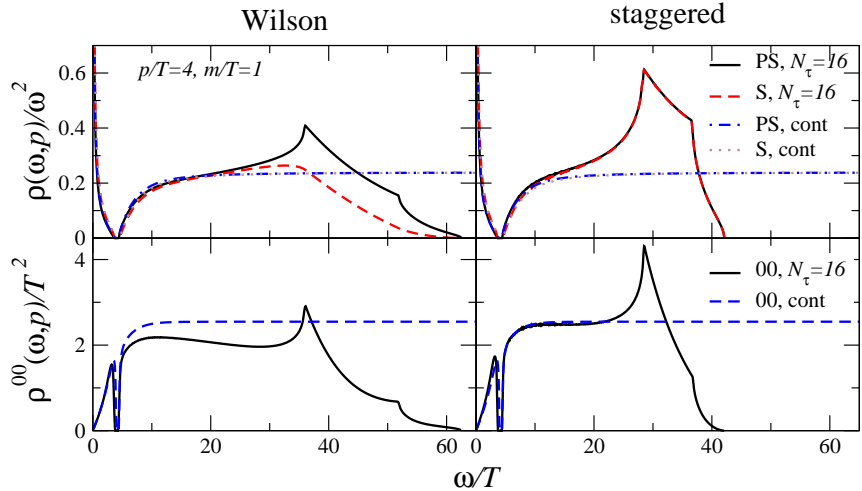}}
 \caption{Pseudoscalar and scalar spectral functions 
 $\rho_{\rm PS,S}(\om,\pv)/\om^2$ (top) and charge density spectral 
 functions $\rho^{00}(\om,\pv)/T^2$ (bottom) as a function of $\om/T$.} 
 \label{fig1}
\end{figure}

We now discuss several characteristics. It is expected from naive 
dimensional arguments that meson spectral functions increase with $\om^2$ 
for large $\om$. This is demonstrated in fig.\ \ref{fig1} (top), where we 
show the (pseudo)scalar spectral functions $\rho_{\rm PS,S}(\om,\pv)$, 
normalized with $\om^2$, in the continuum and on the lattice for finite 
$N_\tau=16$. In order to approximate the thermodynamic limit, $N_\sigma$ 
is taken very large, typically $\sim 2000$ \cite{Aarts:2005hg}. We only 
show results for isotropic lattices here, results for anisotropic lattices 
can be found in ref.\ \cite{Aarts:2005hg}. The cusps at larger energies 
are due to the finite Brillouin zone and are therefore lattice artefacts. 
They are located at $\om = E_{\kv-\pv/2} + E_{\kv+\pv/2} \approx 
2E_{\kv}$, with $\kv = (\pi/a,0,0), (\pi/a,\pi/a,0)$ $+$ permutations for 
Wilson fermions. The maximal energy is determined by $\kv = 
(\pi/a,\pi/a,\pi/a)$ ($\kv = (\pi/2a,\pi/2a,\pi/2a)$ for staggered 
fermions). For Wilson fermions $\rho_{\rm PS}$ and $\rho_{\rm S}$ differ, 
due to the presence of the Wilson mass term, while for staggered and 
continuum fermions they are degenerate for larger energies.

Charge density spectral functions do not increase with $\om^2$, but 
instead reach a constant value: $\rho^{00}(\om,\pv) \to N_cp^2/6\pi$ 
(vector charge density) and $\rho^{00}_{5}(\om,\pv) \to 
N_c(p^2+6m^2)/6\pi$ (axial charge density). Note that this is relevant 
for the choice of default model in the Maximal Entropy Method
\cite{Asakawa:2000tr}. For the 
vector charge density, this large energy behaviour follows from current 
conservation. In fact, at zero momentum current conservation completely 
fixes the charge density spectral function to be $\rho^{00}(\om,\vecnul) = 
2\pi\chi \om\delta(\om)$, where $\chi$ is the charge susceptibility. The 
corresponding euclidean correlator is then constant, 
$G^{00}(\tau,\vecnul)=T\chi$. On the lattice these expressions only hold 
when the conserved current is used, and not the local one. The continuum 
and lattice spectral functions $\rho^{00}$ are compared in fig.\ 
\ref{fig1} (bottom) for nonzero $p=4T$. Note that the local definition is 
used here. The staggered result appears to track the continuum one up to 
larger energies than the one obtained with Wilson fermions.

\begin{figure}[t]
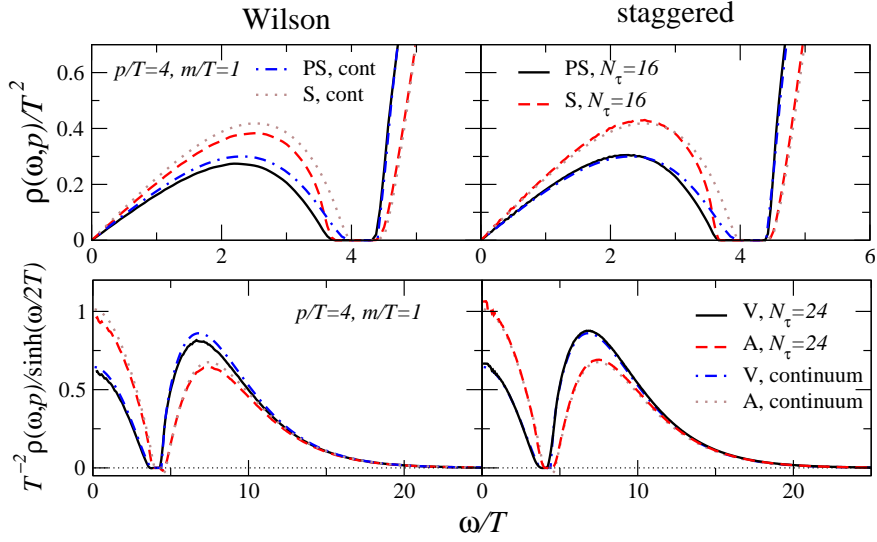

 \centerline{\epsfig{file=plotPSSzoom_mod.eps,height=4cm}}
\vspace{-0.9cm}
 \centerline{\epsfig{file=plotVAsinh_mod.eps,height=4cm}}
 \caption{Low energy region of $\rho_{\rm PS,S}(\om, \pv)$, normalized 
 with $T^2$, from fig.\ 1 (top) and 
 $\rho_{\rm V,A}(\om,\pv)$, normalized with $T^2\sinh(\om/2T)$, (bottom) 
 as a function of $\om/T$.}
 \label{fig2}
\end{figure}

Clearly visible in fig.\ \ref{fig1} is the lightcone is at $\om=p=4T$ and 
the spectral weight below the lightcone. In the (pseudo)scalar channel, a 
better way to view this physically interesting region is by normalizing 
$\rho_{\rm PS,S}(\om,\pv)$ with $T^2$ instead of with $\om^2$. The result 
is shown in fig.~\ref{fig2} (top). The spectral functions increase 
linearly with $\om$ for small $\om$ and vanish at the lightcone. There is 
a gap when $p<\om<\sqrt{p^2+4m^2}$ (higher loop corrections will fill this 
gap). The scalar and pseudoscalar spectral functions differ due to the 
nonzero quark mass. Again the staggered result compares better with the 
continuum one. The main lattice artefact in this region is the mismatch 
between the continuum and the lattice lightcone, which can be quite 
substantial.

\begin{figure}[t]
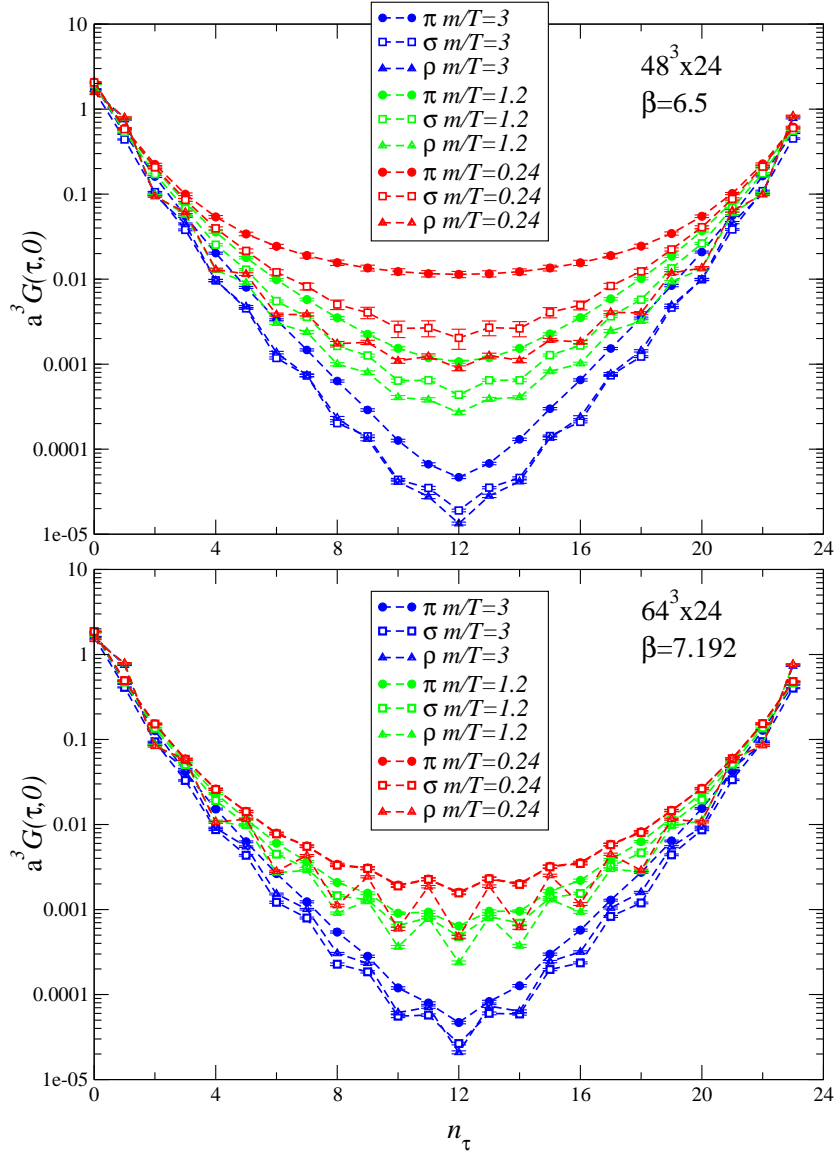

 \centerline{\epsfig{file=48x48x48x24_mod.eps,width=11cm}}
\vspace{-0.cm}
 \centerline{\epsfig{file=64x64x64x24_Z3.eps,width=11cm}}
 \caption{Euclidean correlators in the pseudoscalar ($\pi$), scalar
 ($\sigma$) and local vector ($\rho$, averaged over $i=1,2,3$) channel for
 three values of the staggered fermion mass $m$ (in units of the 
 temperature) in
 quenched QCD on a lattice with $T\approx 160$ MeV (top) and
 $T\approx 420$ MeV (bottom).}
 \label{fig3}
\end{figure}

A very convenient way to present both the low and the high energy 
behaviour of spectral functions in one figure is to show 
$\rho_H(\om,\pv)/\sinh(\om/2T)$, i.e.\ the integrand in eq.\ 
(\ref{eqGrho}) at the midpoint $\tau=1/2T$. This combination takes a 
finite value when $\om\to 0$ and vanishes exponentially for large $\om$. 
We show $\rho_{\rm V} = \rho^{ii}-\rho^{00}$ and $\rho_{\rm A} = 
\rho^{ii}_5-\rho^{00}_5$ in fig.\ \ref{fig2} (bottom). The lattice 
artefacts at large $\om$ are now exponentially suppressed, which implies 
that the euclidean correlator around $\tau=1/2T$ is not very sensitive to 
those. Finally, the area under the two curves is identical, 
which follows from a surprising relation (``sum rule'') at this order in the 
loop expansion \cite{Aarts:2005hg}.

\section{Outlook}

In Fig.\ \ref{fig3} we show preliminary results from a quenched simulation 
with staggered lattice fermions. To facilitate comparison of hadron 
correlators at differing temperatures, the lattice parameters have been 
chosen to reproduce temperatures both below ($\beta=6.5$, $48^3\times24$) 
and above ($\beta=7.192$, $64^3\times24$) the deconfining temperature with 
the same number of temporal spacings $N_\tau=24$. The data shown are for 
zero spatial momentum; comparison of the $\pi$ and $\sigma$-propagators 
between the two sets, particularly for the lightest mass $m/T=0.24$ (shown 
in red), show clear evidence for chiral symmetry restoration in the 
deconfined phase. A spectral function analysis of these data using the 
Maximal Entropy Method is in progress.

\section*{Acknowledgements}
 S.K.\ thanks PPARC for support during his visit to Swansea in 2004/05.

\end{document}